\documentclass[twocolumn,preprintnumbers,amsmath,amssymb,10pt,a4paper,showpacs,superscriptaddress,nofootinbib]{revtex4}
\usepackage{geometry,hyperref}
\usepackage{graphicx}
\usepackage{dcolumn}
\usepackage{bm}
\newcommand{\be}{\begin{equation}}
\newcommand{\ee}{\end{equation}}
\newcommand{\bd}{\begin{displaymath}}
\newcommand{\ed}{\end{displaymath}}
\newcommand{\BE}{\begin{eqnarray}}
\newcommand{\EE}{\end{eqnarray}}

\newcommand{\sm}{\setminus}

\newcommand{\lap}[1]{\nabla^{#1}}

\begin{document} 

\title{Voter models with conserved dynamics}

\author{Fabio Caccioli}
\email{caccioli@santafe.edu}
\affiliation{Santa Fe Institute, Hyde Park Road, Santa Fe, NM, USA}

\author{Luca Dall'Asta}
\email{luca.dallasta@polito.it}
\affiliation{DISAT and Center for Computational Sciences, Politecnico di Torino,
Corso Duca degli Abruzzi 24, 10129 Torino, Italy}
\affiliation{Collegio Carlo Alberto, Via Real Collegio 30, 10024 Moncalieri, Italy}

\author{Tobias Galla}
\email{tobias.galla@manchester.ac.uk}
\affiliation{Theoretical Physics, School of Physics and Astronomy, The University of Manchester, Manchester M139PL, United Kingdom}

\author{Tim Rogers}
\email{tim.rogers@manchester.ac.uk}
\affiliation{Theoretical Physics, School of Physics and Astronomy, The University of Manchester, Manchester M139PL, United Kingdom}

\begin{abstract} We propose a modified voter model with locally conserved magnetization and investigate its phase ordering dynamics in two dimensions in numerical simulations. Imposing a local constraint on the dynamics has the surprising effect of speeding up the phase ordering process. The system is shown to exhibit a scaling regime characterized by algebraic domain growth, at odds with the logarithmic coarsening of the standard voter model.  A phenomenological approach based on cluster diffusion and similar to Smoluchowski ripening correctly predicts the observed scaling regime. Our analysis exposes unexpected complexity in the phase ordering dynamics without thermodynamic potential.
\end{abstract}

\maketitle

How do systems of interacting particles order, and what are the mechanisms by which microscopic interaction generates macroscopic structures? These questions have first been asked in  biology \cite{turing} and then in condensed matter physics, magnetism and statistical mechanics, where kinetic ordering dynamics has received considerable attention  \cite{B94, KRB10}. Ordering processes have more recently also become of interest in interacting agent models of social processes \cite{Schelling, CFL07}. Answering our opening questions requires an off-equilibrium approach starting from the microscopic dynamics itself -- a challenge of considerable intricacy. Our understanding of processes out of equilibrium is still limited, and developing a more complete picture is a focus of current research.  

Processes of kinetic ordering are traditionally addressed in the context of simple lattice models. In condensed matter physics their dynamics is usually based on the minimisation of a thermodynamic potential, coupled to a heat bath and obeying detailed balance. Such models allow for a characterization of different types of ordering. There is a distinction between models with and without a local conservation law. Glauber dynamics and the kinetic Ising model of magnetic systems have no conservation law, their characteristic length scale grows algebraically with time, $L(t) \sim t^n$, where the coarsening exponent is found to take the value $n=1/2$ in phenomenological theories \cite{B94, KRB10}. In the conserved Kawasaki dynamics, on the other hand, one finds $n=1/3$ \cite{B94, KRB10}. This type of dynamics is used to describe alloys or binary liquids. In both models the coarsening dynamics is driven by surface tension.
  
A second class of dynamics is defined by systems lacking a thermodynamic potential or energy function. Such models are frequently motivated by human social interaction \cite{CFL07}. This type of model, frequently based on pair-wise interaction, has also been used to study autocatalytic chemical reactions \cite{ziff} and bacterial populations \cite{frey}. In lattice models with $\mathbb{Z}_2$-symmetry and absorbing states domain coarsening can proceed in the absence of surface tension, driven by interfacial noise \cite{DCCH01,ACDM05}. The main representative of this class is the voter model (VM) \cite{L85}, in which spin-like variables align with the state of a randomly chosen nearest neighbor. This dynamic does not permit a locally conserved quantity, even though global magnetization is conserved in the ensemble. The coarsening dynamics of the VM is algebraic (with $n = 1/2$) in spatial dimension $d=1$, and logarithmic in $d=d_c=2$, that is,  $L(t) \propto 1/\ln{t}$ \cite{K92,FK96}. For $d>d_c$ the interfacial noise becomes irrelevant and the infinite system does not order.

The purpose of our work is to investigate the phase separation dynamics in models of the voter type, but with local conservation of the order parameter. Specifically we study spin-exchange processes based on interactions of pairs of particles. Similar to the celebrated Schelling model of segregation \cite{Schelling} such dynamics can be motivated by social processes. This is not the main objective we wish to pursue though. We aim to systematically work towards a more complete picture of the possible types of ordering dynamics in off-equilibrium  particle models. Specifically we introduce variants of the VM with local conservation laws. While some VM modified along these lines do not order at all, others show an effective algebraic domain growth. We provide a phenomenological understanding of this behaviour invoking a combination of two different mechanisms: (1) a linear instability triggering the formation of compact faceted patterns, and (2) a process similar to Smoluchowski ripening \cite{smoluchowski}, responsible for domain coarsening in the long-time regime. The latter occurs through cluster coalescence driven by surface diffusion and follows an algebraic law $L(t) \sim t^{1/5}$. Interestingly, imposing the local constraint does not slow down the dynamics, but speeds it up relative to the logarithmic ordering of the standard unconstrained VM \cite{schweitzer}.

{\em Microscopic Dynamics.} Consider binary variables $s_i \in \{+1,-1\}$ defined on sites $i$ of a $d$-dimensional lattice. At each time step two neighboring sites $i$ and $j$ are chosen at random. If $s_i\neq s_j$ then the two spins are exchanged with probability
\be
 R^{s_i s_j}_{ij}= \frac{1 - s_i h_{i \sm j}}{2}  \frac{1 - s_j h_{j \sm i}}{2},
\ee
where $h_{i\sm j} = \frac{1}{2d-1}\sum_{k\in \partial i \sm j} s_k$ and where $\partial i$ is the neighborhood of $i$.
These dynamics can be understood as follows: each of the two sites $i,j$ randomly selects one of their remaining neighbors, $k \in \partial i \setminus j$ and $\ell \in \partial j \setminus i$. If $s_\ell = s_i$ and $s_k = s_j$, then $s_i$ and $s_j$ are exchanged.  A more detailed motivation can be found in the accompanying EPAPS document \cite{SI}, as well as a discussion of other variants of the model.   
 
\begin{figure}[t]\includegraphics[width=1.\columnwidth]{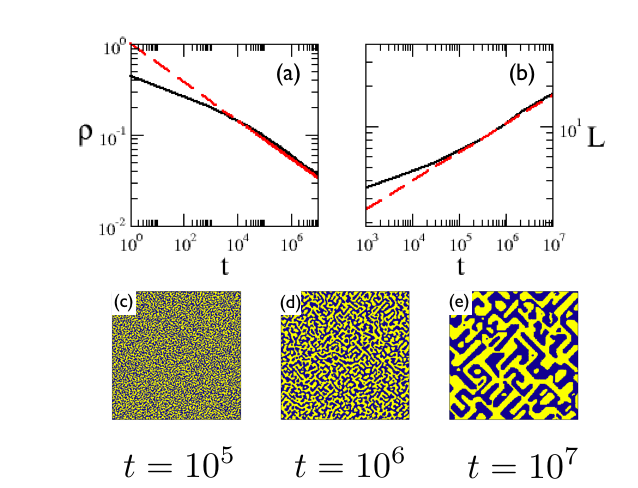}
\caption{\label{fig:fig1}
Voter model with microscopically conserved dynamics. Panels (a) and (b) show the time-evolution of the density of interfaces and correlation length respectively, panels (c) - (e) snapshots of the kinetic ordering process. Dashed lines in (a) and (b) indicate algebraic growth laws with exponent $0.21$. Data is from simulations of a $500\times 500$ system.} 
\end{figure}

We consider unbiased random initial conditions in $d=2$ dimensions. The temporal behavior of the density of interfaces (neighboring pairs of opposing spins)  $\rho(t)$, shown in Fig.~\ref{fig:fig1}(a), reveals a very slow initial dynamical regime, in which clusters of spins with same orientation are formed. At long times the system enters a regime of algebraic coarsening, in which the density of interfaces decays according to $\rho(t)\sim t^{-x}$. This behaviour is similar to that seen in curvature-driven phase separation, and in contrast to the usual VM in $d=2$ which exhibits logarithmic coarsening. In this dynamic scaling regime the system has a single characteristic length scale, $L(t)$, and the order-parameter correlation function assumes a scaling form $C(r,t) = f(r/L(t))$. The typical length scale can be extracted from the first zero of the correlation function, $C(r,t)$ and is found to grow algebraically, $L(t) \propto t^{n}$, see Fig.~\ref{fig:fig1}(b). From our data the coarsening exponent is found as $x\simeq n\simeq 0.21$, and therefore different from standard the exponent $1/3$ in curvature-driven coarsening with locally conserved order-parameter \cite{B94}.  Snapshots from simulations of the conserved VM are shown in Fig.~\ref{fig:fig1}©-(e), with time the system develops faceted domains with straight and elongated diagonal boundaries.

{\em Pattern formation and mean-field approach.}  Unlike standard VM the stochastic dynamics of the present model cannot be solved exactly due to the non-linearity of the transition rates.  Following classical approaches \cite{KRB10}, we neglect correlations between different sites and derive mean-field evolution equations replacing the average of products of spins $\langle s_i \cdots s_j\rangle$ by the product of their averages $\langle s_i \rangle \cdots \langle s_j\rangle$. We will use the variable $\varphi_i$ to identify the mean-field approximation of the average spin value  $\langle s_i \rangle$.
In a continuous-time limit, the evolution equation for the local magnetization $\varphi_i$ becomes 
\be\label{eq:motion}
\dot\varphi_i = \frac{2}{d} \sum_{j\in\partial i}\left[ \frac{1-\varphi_i}{2} \frac{1+\varphi_j}{2} R_{ij}^{-+} -\frac{1+\varphi_i}{2}\frac{1-\varphi_\ell}{2} R_{ij}^{+-}\right],
\ee
resulting in a lengthy expression on the right-hand-side, reported in detail in the accompanying EPAPS document \cite{SI}. These mean-field equations satisfy the conservation law $\sum_i \dot\varphi_i=0$. 
The mean-field description can be used to explain the process of pattern formation. Linearizing \eqref{eq:motion} around the homogeneous solution $\varphi_i \equiv 0$ $\forall i$, one finds 
\begin{equation}\label{eq:motionlin} 
\dot \varphi_i \propto  - \left[ \lap{4} + \frac{1}{d}\lap{2}\right]\varphi_i, 
\end{equation}
where $\lap{2}\varphi_i$ is the (lattice) Laplacian.
The present model is therefore characterized by a linear instability of type II in the Cross-Hohenberg classification \cite{CG09} with a band of unstable wave numbers $0 < k <1/\sqrt{d}$ with maximum growth rate at $k_{\mbox{\tiny max}} = 1/\sqrt{2d}$. A linear instability generated by terms of the type as in \eqref{eq:motionlin} is typical of coarsening with conserved dynamics, described by the Cahn-Hilliard equation \cite{B94}. These terms have a different origin in our model though. In the Cahn-Hilliard equation the term proportional to $-\lap{2} \varphi$ comes from linearizing the thermodynamic potential; in the present model it is instead generated dynamically by exclusion of spin $j$ from the sampling of $i$'s neighborhood and $i$ from the sampling of $j$'s neighborhood. This restriction prevents {\em bulk diffusion}, i.e. the migration of isolated defects, which is present in the Kawasaki spin-exchange dynamics even at $T=0$ \cite{KRB10}. The influence of the restriction becomes progressively less important in higher dimensions, and the linear instability is expected to disappear for $d \to \infty$. If we remove the site restriction, only the surface diffusion term, $-\nabla^4\varphi$, survives in Eq. \eqref{eq:motionlin}, and phase separation does not occur \cite{KRB10, SI}. 
\begin{figure}[t]
\vspace{-3em}
\includegraphics[width=1\columnwidth]{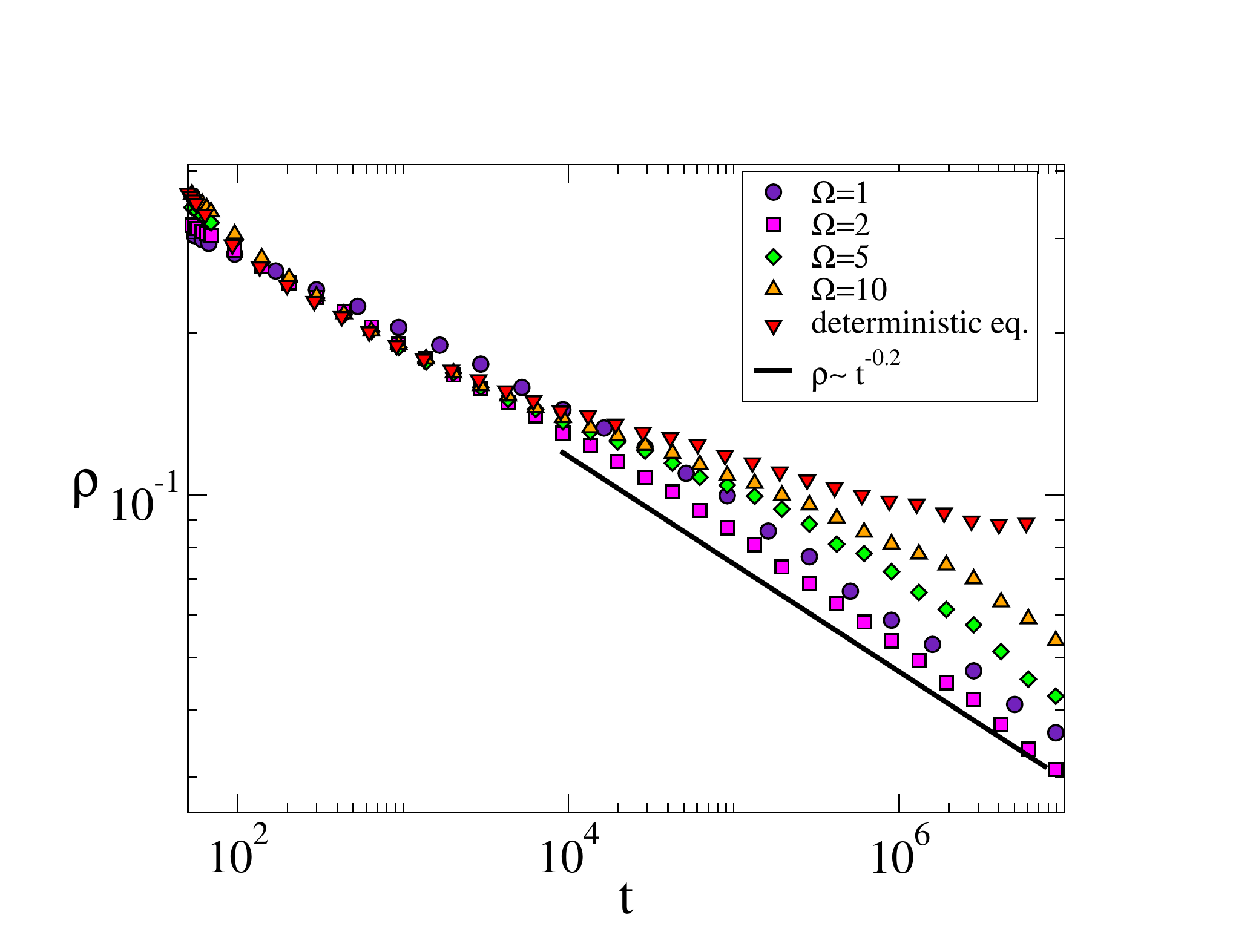}
\caption{(Color online). Interface density from a numerical integration of the mean-field dynamics, Eq. (\ref{eq:motion}), and from the microscopic model with $\Omega$ spins per site. Data is from single simulation runs of a system of size $200\times 200$.} 
\label{fig2}
\end{figure}

Although the mean-field approach describes the pattern formation process qualitatively to some extent, it does not capture the correct long-time coarsening regime. This is not surprising, in the example of the standard VM the mean-field dynamics is simply given by diffusion equation, and it does not capture the essential features of the dynamics in low spatial dimensions (e.g. $d=2$), where the fluctuations due to multiplicative interfacial noise are relevant.  The conserved VM we discuss here is more complicated, nevertheless the differences between the dynamics of the original microscopic model and the mean-field equations can again be traced back to the discreteness of the degrees of freedom, i.e. to multiplicative noise. We have considered a modified model in which each lattice site contains $\Omega$ spins, similar to what was proposed in \cite{oursblythe}. In a given update step a spin interacts with randomly chosen spins in the neighboring sites, following the microscopic rules of the conserved VM. For $\Omega \to \infty$, the dynamics is deterministic and given by the above mean-field equations. The data in Fig.~\ref{fig2} show that for $\Omega$ large but finite, the temporal behavior is very similar to the predictions of the mean-field equations in an initial brief coarsening regime governed by surface diffusion (also referred to as `bidiffusion'  \cite{KRB10}) and with algebraic coarsening with $n \approx 1/4$. Then the pattern formation process takes place and the dynamics finally reaches a structured configuration of patterns, with only a slow further coarsening process or potentially none. Considering the case of small $\Omega\approx 1$, instead, the system develops an algebraic coarsening regime with exponent $n \simeq 0.2$, which persists in the long-time regime.
 
\begin{figure}[t]
\includegraphics[width=0.75\columnwidth]{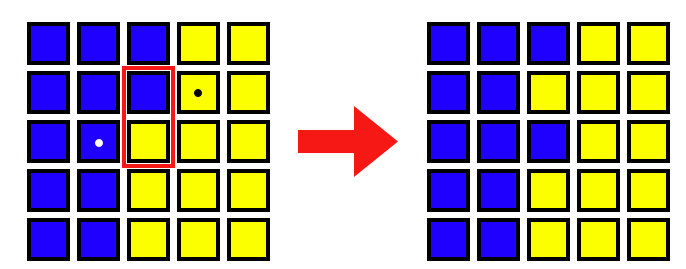}
\caption{(Color online) Periphery diffusion: The exchange process of the two spins marked by the red (grey) frame, along with their respective neighbors (marked by dots), leads to a rearrangement of kinks at the interface between two clusters. This may result in a diffusion dynamics of the centre of mass of clusters (cluster diffusion).} 
\label{fig3}
\end{figure}

{\em Coarsening by Cluster Diffusion.} At this point it is clear that the observed asymptotic coarsening regime is an intrinsic effect of individual discreteness, and that it cannot be captured by simple mean-field equations.  Instead we follow a phenomenological approach, relating to a phenomenon known in material science as Smoluchowski ripening \cite{smoluchowski}, to obtain further insight into the process of cluster aggregation. In absence of bulk diffusion, we can exclude Ostwald ripening \cite{ostwald}, therefore we expect a slow coarsening process, with $n < 1/3$. Horizontal and vertical domain walls are blocked, while other interfaces can still move by means of a process known as `periphery diffusion', that is a slow diffusional movement of kinks on the boundary of a domain (see the illustration in Fig.~\ref{fig3}). This movement causes a slow effective displacement of the center of mass of the clusters, i.e. an effective {\em cluster diffusion} process. Larger clusters move more slowly because the effective diffusion coefficient of the cluster depends on the ratio between the surface and the volume of the clusters. The form of the resulting diffusion coefficient of clusters has been derived in the literature, see e.g. \cite{siclen}. In order to move the center of mass of a cluster by a distance $\delta_c$ in two dimensions, we need order $L^2$ spins to move by the same distance. This corresponds to a single spin moving over a distance $\delta_p \propto L^2 \delta_c$. On the other hand, the rate with which moves of the centre of mass of a cluster occur is proportional to that of diffusion events on its surface, i.e. $\Gamma_c \propto \Gamma_p L$. Given that the single spin diffusion coefficient is $D_p \propto \delta_p^2 \Gamma_p$, we obtain that the diffusion coefficient of a cluster of typical scale $L$ decays as $D_c(L)  \propto L^{-3}$.   
\begin{figure}[t]
\vspace{-3em}
\includegraphics[width=1\columnwidth]{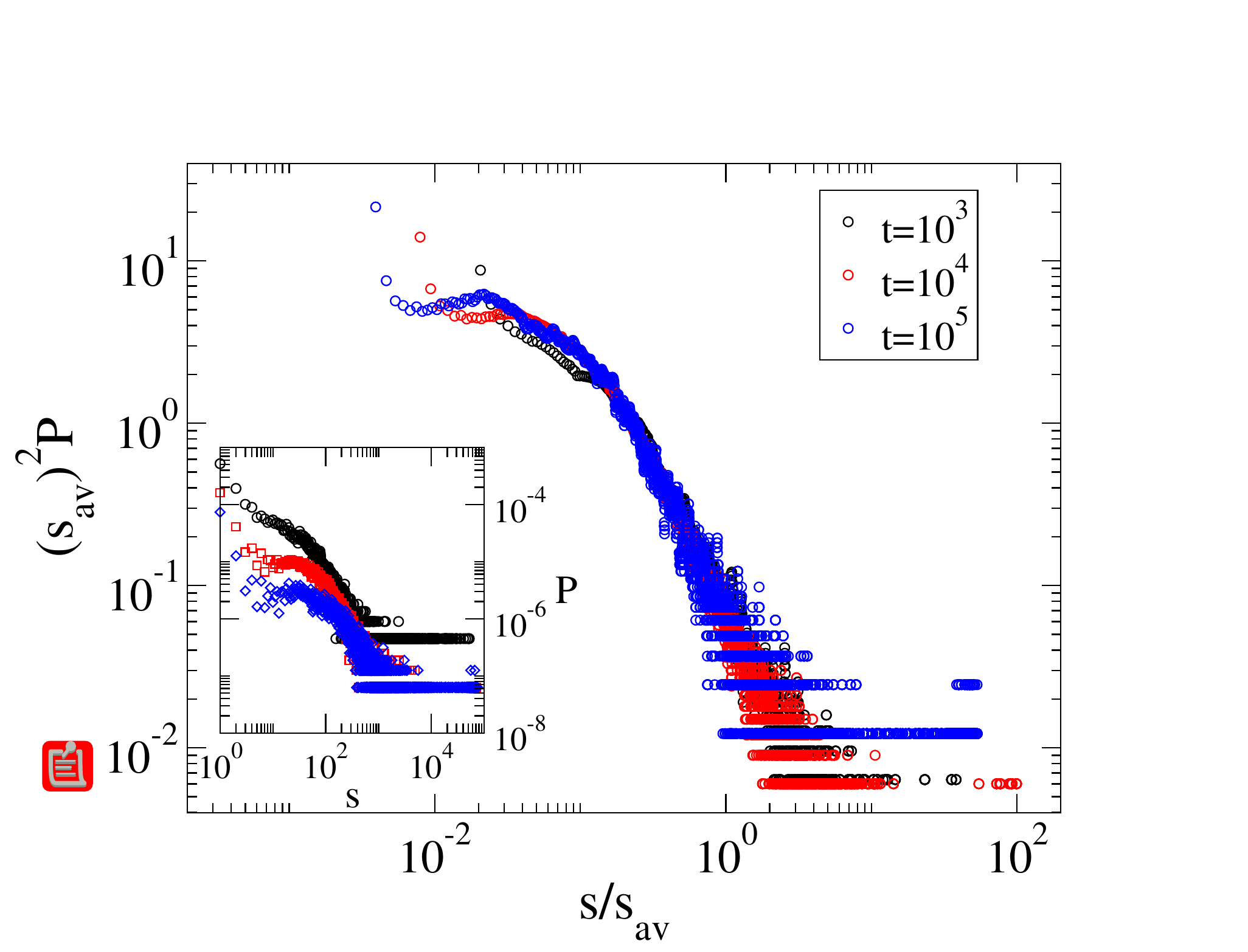}
\caption{(Color online). Scaling behavior of the cluster-size distribution $P(s,t)$. Main panel shows the collapse plot obtained using the predicted scaling relation Eq. (\ref{eq:scaling}), the inset shows the raw data for $P(s,t)$. Data is from $10$ runs of a system of size $400\times 400$ (running average performed for data in main panel, see also \cite{SI}).} 
\label{fig4}
\end{figure}
In order to find the coarsening exponent we next consider the temporal evolution of the average cluster size by means of the Smoluchowski equation for cluster coagulation processes. This mean-field approach is usually considered qualitatively correct also in two dimensions, Kang and Redner \cite{KR84} have shown that the upper critical dimension is $d_c=2$. We therefore expect logarithmic corrections at most. We follow the derivation proposed by Kandel \cite{K97}, and introduce the size (area) of a cluster in $d$ dimensions, $s$. The time-dependent density of clusters of size $s$ per lattice site is denoted by $P(s,t)$ \cite{remarkp}. 
One then finds the following scaling relation  in the long-time limit \cite{K97}
\be\label{eq:scaling}
P(s,t) = t^{-\alpha} f\left[s/s_{\mbox{\footnotesize av}}(t)\right],
\ee
where the average cluster size $s_{\mbox{\footnotesize av}}(t)=\sum_s sP(s,t)/\sum_s P(s,t)$ scales as $s_{\mbox{\footnotesize av}}(t) \sim t^{\beta}$. The scaling exponents are predicted to be  $\alpha = 2/(\zeta+1)$ and $\beta = 1/(\zeta+1)$, where $\zeta$ characterizes the scaling of the diffusion constant of clusters, with the cluster size, $D_c(s)\sim s^{-\zeta}$ \cite{K97}. The scaling of Eq. (\ref{eq:scaling}) is confirmed by simulations of the microscopic model, see Fig.~\ref{fig4}, adding weight to the hypothesis that a process similar to Smoluchowski ripening is at work in our system. 
In the present case we have $D_c(L)\sim L^{-3}$, i.e. $\zeta = 3/2$. Therefore we expect that the asymptotic behavior of the microscopic model is characterized by a coarsening regime $s_{\mbox{\footnotesize av}} \sim t^{2/5}$. Given that $L$ is the unique length scale the hypothesis of Smoluchowski ripening leads to $L\sim t^n$ with $n = \beta/d = 1/5$. The data shown in Fig.~\ref{fig:fig1} is consistent with this phenomenological prediction.  
We note though that the facetted configurations shown in Fig.~\ref{fig:fig1} differ from those seen in physical systems undergoing Smoluchowski ripening \cite{thiel}. In these systems one species of particles is frequently more abundant than the other, which may explain the difference in the resulting structures \cite{remark}. 

{\em Globally conserved dynamics.} To determine if the modified scaling behaviour we observe is genuinely an effect of local (microscopic) conservation, as opposed to conservation \textit{per se}, we considered a version of the dynamics with conservation only at a global level \cite{SI}. We find logarithmic coarsening as in the standard VM \cite{SI}, in-line with earlier work indicating that conservation laws at long-range do not affect the coarsening dynamics of non-conserved models \cite{B93}.

{\em Conclusions.} We have put forward a voter-like model with strict microscopic conservation of magnetization. The spin-exchange rule introduces non-linearities that turn the logarithmic growth of the standard VM into an algebraic coarsening process, driven by diffusion at the surfaces of clusters and by coalescence of clusters, similar to what is known as Smoluchowski ripening. Simulation results are consistent with a growth law $L\sim t^{1/5}$, as predicted by a phenomenological theory. These findings extend the complexity and variety of the ordering of interacting particle models out of equilibrium. In particular our analysis exposes unexpected behavior in models with conserved dynamics lacking a thermodynamic potential. This, we think, can be relevant not only for non-equilibrium statistical physics and biological processes, but also for models of segregation in the social sciences.

\vspace{1em}
{\em Acknowledgements:} TG is supported by a RCUK (EP/E500048/1). TG and TR acknowledge support by EPSRC (EP/I005765/1 and EP/H02171X/1). LDA acknowledges support by EU FET Open 265496 and FIRB Project RBFR10QUW4. We would like to thank the Eyjafjallaj\"okull volcano for its eruption in 2010, resulting in a unexpected stay of LDA in Manchester, during which this work was initiated. 
\vspace{1em}

\appendix


\begin{thebibliography}{99}
\bibitem{remarkp} $P(s,t)$ is the number of clusters at time $t$ containing precisely $s$ spins, divided by the total number of spins in the system. We have $\sum_s sP(s,t)=1$.
\bibitem{remark} Biasing initial condition in our model we find a very slow coarsening dynamics -- clusters do not appear to coalesce for a significant time. It is therefore hard to assess the asymptotic regime in simulations \cite{SI}.
\end{thebibliography}
\end{document}